\documentclass[preprint,aps,secnumarabic%
,showkeys,amsmath,amssymb]{revtex4}

\usepackage{docs}
\usepackage{natbib}
\usepackage{graphicx}
\usepackage{dcolumn}
\usepackage{bm}

\begin{document}

\title{Liquid Crystal Pretilt Control by Inhomogeneous Surfaces}

\author{Jones T. K. Wan$^{1,\ast}$, Ophelia K. C. Tsui$^{1,}$\footnote{Electronic mails: jwan@ust.hk, phtsui@ust.hk}, Hoi-Sing Kwok$^2$, Ping Sheng$^1$}

\affiliation{$^1$Department of Physics and Institute of Nano
Science and Technology, Hong Kong University of Science and
Technology, Clear Water Bay, Kowloon, Hong Kong.}

\affiliation{$^2$Department of Electrical and Electronic
Engineering, Hong Kong University of Science and Technology, Clear
Water Bay, Kowloon, Hong Kong.}

\received{~~~~~~~~~~~~~~~~~~~~}

\begin{abstract}
We consider the pretilt alignment of nematic liquid crystal (LC)
on inhomogeneous surface patterns comprising patches of
homeotropic or homogeneous alignment domains, with azimuthal
anisotropy assumed in the surface plane. We found that the
resultant LC pretilt generally increases continuously from the
homogeneous limit to the homeotropic limit as the area fraction of
the homeotropic region increases from 0 to 1. The variations are
qualitatively different depending on how the distance between
patches compares to the extrapolation length of the stronger
anchoring domain. Our results agree with those previously found in
stripped patterns. The present findings may provide useful
guidelines for designing inhomogeneous alignment surfaces for
variable LC pretilt control - a subject of much technological
interest in recent years.
\end{abstract}

\maketitle

Study of inhomogeneous surfaces for liquid crystal (LC) alignment
has witnessed a rapid growth in recently years
\cite{guyon_1973,berreman_1979,barbero_1992,qian_prl_1996,ong_jap_1984,Clark_Science_2001,Kim_Nature_2002,
zhang_prl_2003,tsui_pre_2004,lee_apl_2004}. Experiments showed
that these surfaces could have unique applications as in the
making of multi-stable display devices \cite{Kim_Nature_2002} and
the control of LC pretilt alignment
\cite{ong_jap_1984,lee_apl_2004}. This type of alignment surfaces
generally comprise two kinds of domains favoring different LC
orientations. Majority of theoretical treatments had dealt with
domains arranged in alternating stripped \cite{guyon_1973,
berreman_1979,barbero_1992,qian_prl_1996,Clark_Science_2001} or
checkerboard \cite{Kim_Nature_2002,zhang_prl_2003,tsui_pre_2004}
patterns. No systematic calculation has been carried for a patchy
pattern, which is more likely achieved by non-lithographic
techniques that are less costly \cite{ong_jap_1984}. In this
letter, we perform an extensive simulation study on the alignment
of nematic LC on surfaces containing patches of homogeneous or
homeotropic alignment domains, with one embedded in the matrix of the other.\\
    \indent Figure~\ref{pattern} shows the inhomogeneous alignment
surfaces we studied for different area fractions of the
homeotropic region, $p$.
%
%
We consider a cubic simulation cell with sides of unity filled
with nematic LC. The lower ($z=0$) surface is the patterned
surface as shown in Fig.~\ref{pattern}, and the upper surface is
free. The cell is assumed to repeat in the $x$ and $y$ directions.
If each type of domain has surface energy $W_i(\theta,\phi)$
$(i=1,2)$, the unit-area surface energy, $F_s$, would be:
\begin{equation}
F_s(\theta(\bm r),\phi(\bm r)) = \delta(z) \sum_{i=1}^2 f_i
(x,y)W_{i}(\theta({\bm r}),\phi({\bm r})). \label{multi_anchor}
\end{equation}
Here, ${\bm r} = (x,y,z)$ is a position vector in the simulation
cell; the LC director field is denoted by ${\bm n}({\bm
r})\equiv(\cos\theta\cos\phi, \cos\theta\sin\phi, \sin\theta)$,
where $\theta({\bm r})$ and $\phi({\bm r})$ are, respectively, the
zenithal (measured from the surface instead of the surface normal)
and azimuthal angles of the LC director; $f_i (x,y)$ is one if
$(x,y)$ belongs to domain $i$ and zero otherwise so that the area
fraction of domain $i$ is obtainable by integrating $f_i(x,y)$
over the $z=0$ unit surface. Since our focus is in the LC zenithal
(pretilt) alignment produced by the inhomogeneous surface, we
impose anisotropy in the $x$-$y$ plane, i.e., both domains prefer
the same azimuthal orientation along, say, $\phi_0$. We further
assume the local preferred zenithal alignment of domain $1$,
$\theta_{01}=88^\circ$ and that of domian $2$,
$\theta_{02}=6^\circ$ to lift the foreseen degeneracy between the
$\phi(\bm{r})=\phi_0$ and $\phi(\bm{r})=\phi_0 + 180^\circ$
states. With these assumptions, the azimuthal orientation of the
LC is constant in space and equal to $\phi_0$. Hence the surface
energy depends only on $\theta(\bm{r})$. We use the
Rapini-Papoular form of the LC surface energy
\cite{zhang_prl_2003}, $W_i(\theta,\phi_0)=\frac{1}{2}W_{\theta
i}\sin^2(\theta-\theta_{0i})$ where $W_{\theta i}$ is the polar
anchoring strength of domain $i$. For the LC elastic energy
density, $F_{e}$, we assume the the Frank-Oseen form:
\begin{equation}
F_{e}(\bm{n})= \frac{1}{2}\{K_{11}(\nabla\cdot\bm{n})^2
+K_{22}(\bm{n}\cdot\nabla\times\bm{n})^2+K_{33}(\bm{n}\times\nabla\times\bm{n})^2\}.
\label{eq:F-O}
\end{equation}
\noindent Here $K_{11}$, $K_{22}$ and $K_{33}$ are the splay,
twist and bend elastic constants, respectively, for which the
values of the nematic 5CB are used in
simulations\cite{zhang_prl_2003}: $K_{11}=7.0\times 10^{-12}$N,
$K_{22}=3.5\times 10^{-12}$N, $K_{33}=1.0 \times 10^{-11}$N. If
the size of the cubic cell were scaled to $\lambda^3$, the total
energy per unit area is given by
\begin{eqnarray}
\frac{F_\mathrm{tot}(\bm{n})}{\lambda^2} &=& \frac{K_{11}}{\lambda} {\bigg \{} \int_0^1\int_0^1\int_0^1 \tilde{F}_{e}({\bm n}) dxdydz\nonumber \\
    &+& \sum_{i=1}^2 \int_0^1\int_0^1 \frac{\lambda}{\ell_{e i}}f_i (x,y)\frac{\sin^2(\theta - \theta_{0i})}{2} dxdy {\bigg \}},
\label{total_en}
\end{eqnarray}
where $\tilde{F}_{e}\equiv F_{e}/K_{11}$, $\ell_{e i}\equiv
K_{11}/W_{\theta i}$ is the extrapolation length of the domain
$i$. Equation (\ref{total_en}) shows that for the same LC
material, the energy density of the system generally increases
with $(1/\lambda)$, but the critical parameter determining the LC
configuration is the ratio $\lambda/\ell_{e}$ (where $\ell_e$ is
the smaller of $\ell_{ei}$): If $\lambda/\ell_e\ll 1$, the elastic
energy would dominate the surface energy, and vice versa in the
opposite limit. For typical alignment surfaces, the polar
anchoring energy is $10^{-4}-10^{-3}$~Jm$^{-2}$ and hence $\ell_e
\approx 0.01 - 0.1~\mu$m.
We determine the equilibrium LC director field by minimizing
Eq.~(\ref{total_en}) with respect to $\bm{n}(\bm{r})$ numerically
as in Ref.~\cite{tsui_pre_2004}. \\
   \indent We first discuss results obtained in the
$\lambda/\ell_e\ll 1$ limit. Shown in
Fig.~\ref{pretilt_vs_area_analtyic} are the averaged pretilt angle
of the calculated LC alignment at $z=0$, $\theta_{\rm av}(0)$,
plotted versus $p$ for $\lambda/\ell_{e2}= 0.01$ and various
ratios of $W_{\theta 1}/W_{\theta 2}$ from 0.5 to 5.
%
%
As seen, all the curves in Fig.~\ref{pretilt_vs_area_analtyic}
bear similar shape with $\theta_{\rm av}(0)$ demonstrating a
gentle rise from $6^\circ$ as $p$ increases from 0 initially,
followed by an abrupt transition from small to high pretilt
whereupon the rise with $p$ becomes gradual again. For $W_{\theta
1}/W_{\theta 2}=1$, the transition occurs at $p=1$, but shifts to
smaller (larger) $p$ when $W_{\theta
1}/W_{\theta 2}$ increases (decreases).\\
\indent We plot $\sigma^{xy}_\theta(z)$, the standard deviation of
$\theta(\bm{r})$ over the $x$-$y$ plane at $z$, against $z$ for
various $\lambda/\ell_{e2}$ from $0.01$ to $1000$
Fig.~\ref{sd_vs_depth}.
%
%
In these plots, $p=0.125$ and $W_{\theta 1}=W_{\theta 2}$ though
we found that the results are qualitatively independent of the set
values of these parameters. As seen, when $\lambda/\ell_{e2} \ll
1$ $(=0.01, 0.1)$, variations of the LC pretilt in the $x$-$y$
plane is negligibly small for all $z$, suggesting the director
field to be uniform in the present limit. Hence the behaviors
shown in Fig.~\ref{pretilt_vs_area_analtyic} for LC alignment at
$z=0$ are essentially the same as those in the bulk. But as
$\lambda/\ell_{e2}$ increases towards 1, $\sigma^{xy}_\theta(0)$
increases notably, and steadily approaches $\sim 27^\circ$ as
$\lambda/\ell_{e2}$ continues to increase towards 1000. We
postpone discussion of the noted asymptotic value of
$\sigma^{xy}_\theta(0)$ until later. Despite of the large
variations in $\sigma^{xy}_\theta(0)$, all curves display similar
decays towards 0 within $z=\lambda$. Least-square fittings reveal
that the data of Fig.~\ref{sd_vs_depth} can be described very well
by exponential decays with decay lengths $\sim \lambda/2\pi$,
consistent with previous results found in stripped
surface patterns possessing zenithal inhomogeneity \cite{barbero_1992}.\\
   \indent We return to the data of Fig.~\ref{pretilt_vs_area_analtyic}.
In adopting an uniform LC director field, the system reduces the
elastic energy in the expense of a larger surface energy. The
resulting pretilt angle should then equal $\theta_{\rm min}$, the
pretilt angle that minimizes the surface energy term of
Eq.~(\ref{total_en}). It is straightforward to show that
$\theta_{\rm min}$ satisfies:
\begin{equation}
\frac{1}{p}=1-
 \frac{W_{\theta 1} \sin2(\theta_{\rm
min}-\theta_{01})} {W_{\theta 2}\sin2(\theta_{\rm
min}-\theta_{02})}. \label{weak_analytic}
\end{equation}
In Fig.~\ref{pretilt_vs_area_analtyic} we plot $\theta_{\rm min}$
deduced from Eq.~\ref{weak_analytic} versus $p$ for the same set
of $W_{\theta 1}/W_{\theta 2}$ indicated in the figure (dashed
lines). As seen, the data of $\theta_{\rm min}$ demonstrates
excellent agreement with those of $\theta_{\rm av}(0)$ obtained
from simulations, verifying our interpretation that the uniform LC
alignment is due to
$\theta_{\rm min}$.\\
  \indent Next we consider the opposite limit where $\lambda/\ell_e
\gg 1$. In this case, the director field on the surface is no
longer uniform. But as Fig.~\ref{sd_vs_depth} shows, it comprises
large surface-induced deformations that decays
%
%
to zero within a distance of $\sim \lambda/2 \pi$ from the
surface. Hence the homogenized LC alignment in the bulk could be
obtained from that at $z=\lambda$. Shown in
Fig.~{\ref{pretilt_vs_area_lw_1000}} are plots of $\theta_{\rm
av}(\lambda)$ versus $p$ for $\lambda/\ell_{e2} = 1000$ and the
same set of $W_{\theta 1}/W_{\theta 2}$ investigated in
Fig.~\ref{pretilt_vs_area_analtyic}. In the same figure are drawn
the corresponding data at $z=0$ for comparison. As one could see,
the average LC pretilts at both $z=0$ and $z=\lambda$ are
independent of $W_{\theta 1}/W_{\theta 2}$. It is interesting to
note that the $z=0$ data demonstrate a simple linear rise with $p$
from $6^\circ$ to $88^\circ$. This is because when $\lambda/\ell_e
\gg 1$, the surface energy dominates the elastic energy in the LC
system. One may thus assume that the LC director to exactly follow
that of the local preferred direction of the anchoring surface,
and hence $\theta_{\rm av}(0)$ should just be the algebraic
average of local preferred orientation over the surface, i.e.,
$p\theta_{01}+(1-p)\theta_{02}$. If $p=0.125$ as for the cases
exemplified in Fig.~\ref{sd_vs_depth}, $\theta_{\rm
av}(0)=16.25^\circ$. Then the standard deviation of $\theta(x,y)$
at $z=0$ in the present limit would be $27.1^\circ$, in excellent
accord with the value shown in Fig.~\ref{sd_vs_depth} for
$\lambda/\ell_2 = 1000$. The slightly larger value of $\theta_{\rm
av}$ in the bulk than that at $z=0$ is attributable to the assumed
$K_{33}$ being larger than $K_{11}$, which favors less bending in
expense of more splay deformations of the LC director field in
homogenizing into the bulk \cite{berreman_1979}.\\
   \indent In this letter, we have considered in detail the pretilt
alignment of nematic LC on a periodic patchy surface pattern
comprising domains favoring either $88^\circ$ or $6^\circ$
zenithal alignments. If the pattern period is much smaller than
the extrapolation length of the stronger anchoring domain, the LC
alignment is uniform with the LC pretilt being one that minimizes
the surface anchoring energy over one pattern period. But if the
pattern period is much bigger than the extrapolation length of the
stronger domain, the LC alignment at the interface copies the
inhomogeneity of the surface pattern, which, however, dies out
within one pattern period from the surface. The resulting
homogenized LC zenithal alignment in the bulk is approximately
given by the average of the local preferred directions of the
domains weighted by their area fractions in the pattern.
Previously Ong et al. \cite{ong_jap_1984} created random
inhomogeneous surfaces by depositing a discontinuous metal film on
a silane (homeotropic agent) coated corrugated SiO$_2$
(homogeneous) substrate, then subjecting the sample to glow
discharge to remove the silane unprotected by the metal islands,
before dissolving the metal in the end. The average LC pretilt on
these surfaces was found to increase monotonically from $0^\circ$
to $90^\circ$ as the average metal film thickness was increased,
demonstrating good qualitative agreement with our results.
Our results also display good consistency with those of Barbero et
al. \cite{barbero_1992} who obtained analytic solutions for LC
alignment on stripped patterns favoring alternatively two
dissimilar zenithal orientations (though the two preferred
orientations must be sufficiently similar
to validate their calculations). As most discussions presented
here are general to LC alignment on inhomogeneous surfaces, it is
probably that our results could be applicable to more general
inhomogeneous patterns with irregular shapes. It is hoped that our
study would motivate more experiments along the
same line of thought.\\
\indent We acknowledge support of the Research Grant Council of
Hong Kong through project no. HKUST6115/03E.

\clearpage
\newpage
%
\begin{figure}[h]
\caption{Patterns used to simulate the inhomogeneous surface.
Filled squares stand for the homeotropic domains; unfilled squares
stand for the homogeneous domains. Periodic boundary conditions
are imposed in the $x$ and $y$ directions.} \label{pattern}
\end{figure}
%
%
\begin{figure}[h]
\caption{Simulation results of average pretilt angle $\theta_{\rm
av}(z=0)$ versus $p$ for $\lambda/\ell_{e2} = 0.01$ and various
$W_{\theta 1}/W_{\theta 2}$ = 0.5 (crosses), 1.0 (triangles), 2.0
(diamonds) and 5.0 (squares) are shown. Dashed lines are the
simulated $\theta_{\rm min}$ versus $p$ according to
Eq.~(\ref{weak_analytic}) for the same set of $W_{\theta
1}/W_{\theta 2}$.} \label{pretilt_vs_area_analtyic}
\end{figure}
%
%
\begin{figure}[h]
\caption{Profiles of the standard deviation of the LC pretilt
along $z$ for different ratios of $\lambda/\ell_{e2}$ from 0.01 to
1000. In these data, $p=0.125$ and $W_{\theta 1}=W_{\theta 2}$.}
\label{sd_vs_depth}
\end{figure}
%
%
\begin{figure}[h]
\caption{The average LC pretilt at $z = \lambda$, which is
essentially the bulk value, versus $p$ for
$(\lambda/\ell_{e2}=1000)$ for four different $W_{\theta
1}/W_{\theta 2}$ as indicated. The data for average pretilt at
$z=0$ are also drawn (solid line).}
\label{pretilt_vs_area_lw_1000}
\end{figure}
%

\newpage
\includegraphics[height=\columnwidth,angle=270]{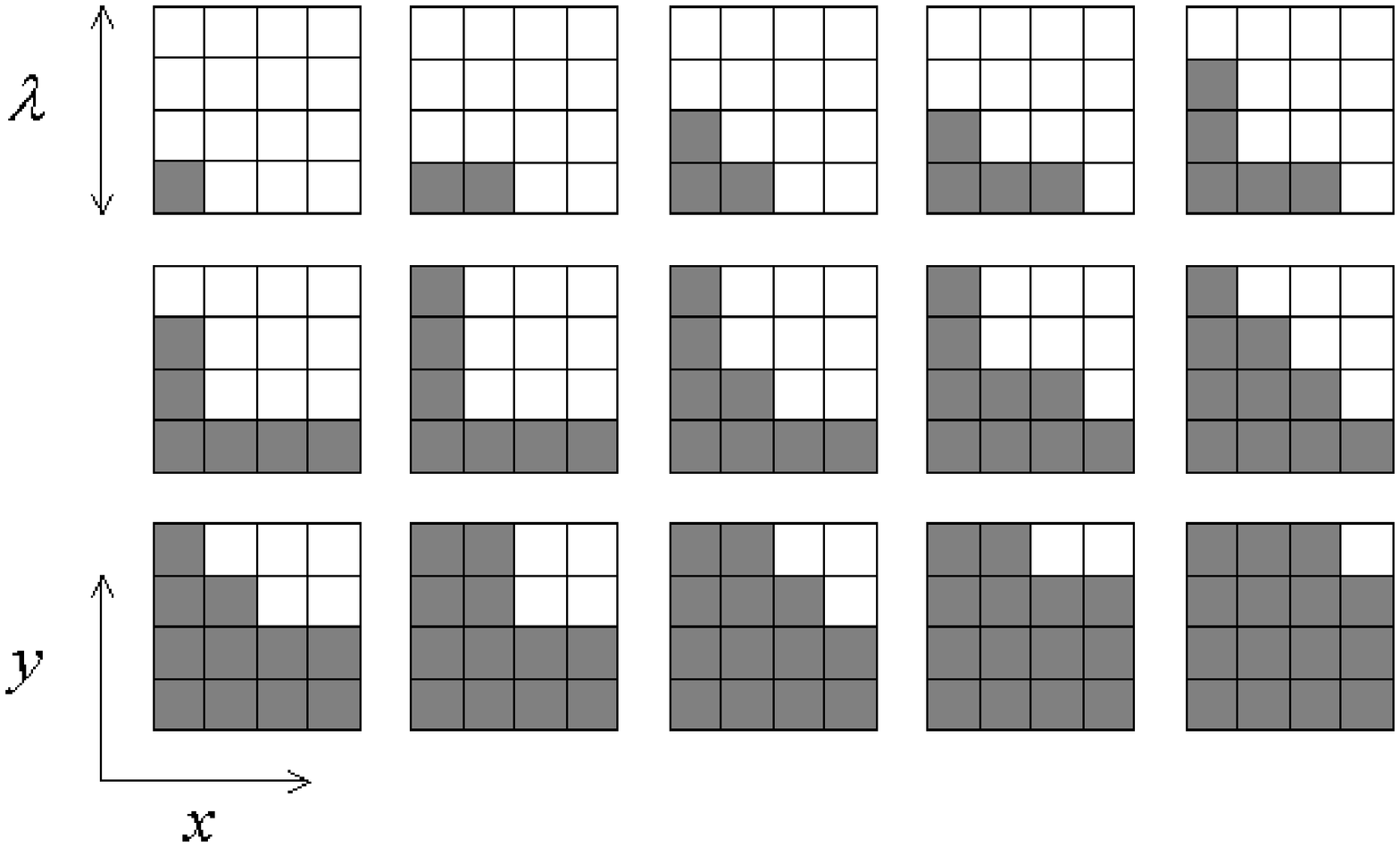}
\centerline{Fig.~\ref{pattern} by Wan et al.}

\newpage
\includegraphics[height=\columnwidth,angle=270]{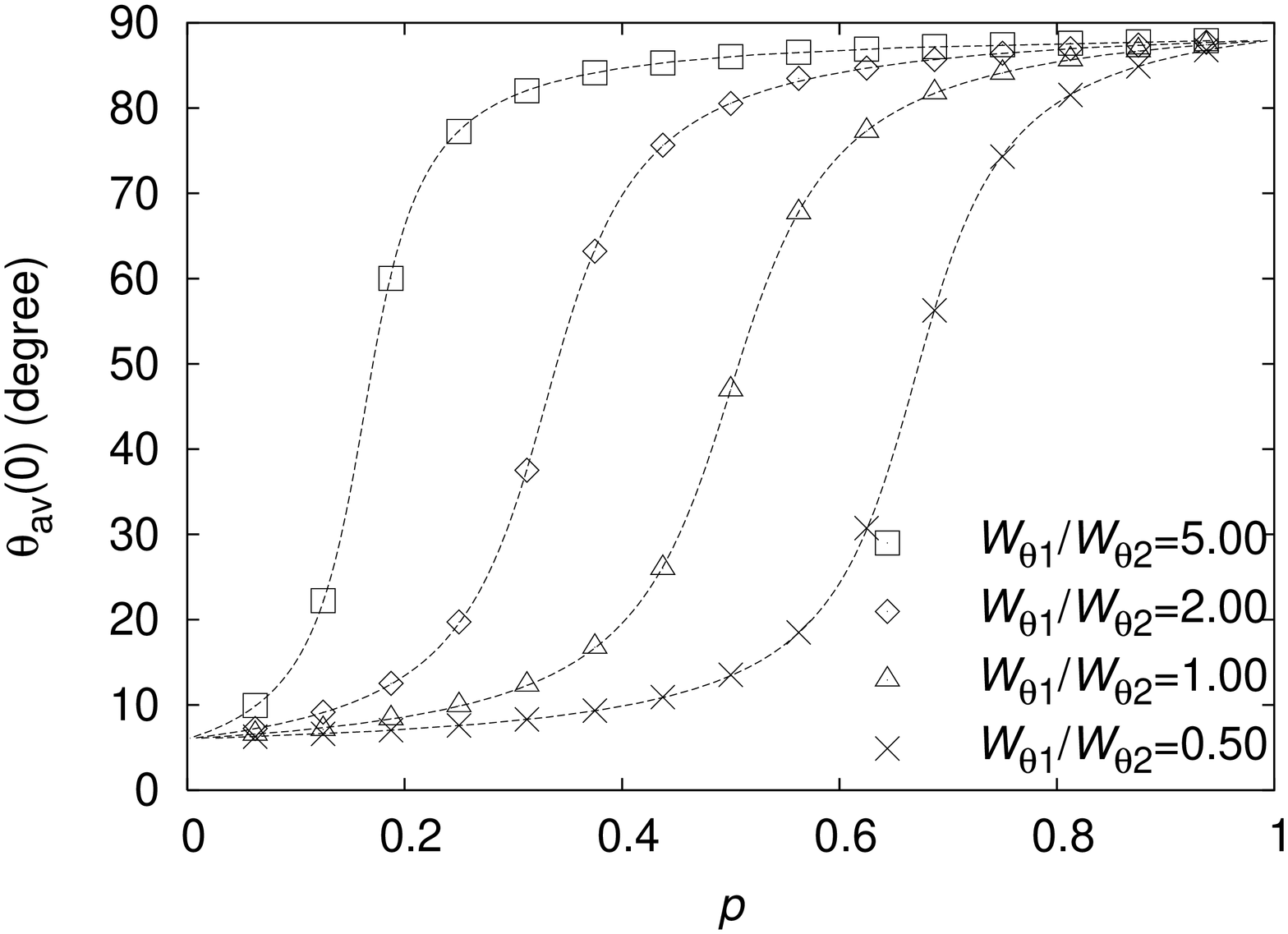}
\\
\\
\\
\\
\centerline{Fig.~\ref{pretilt_vs_area_analtyic} by Wan et al.}

\newpage
\includegraphics[height=\columnwidth,angle=270]{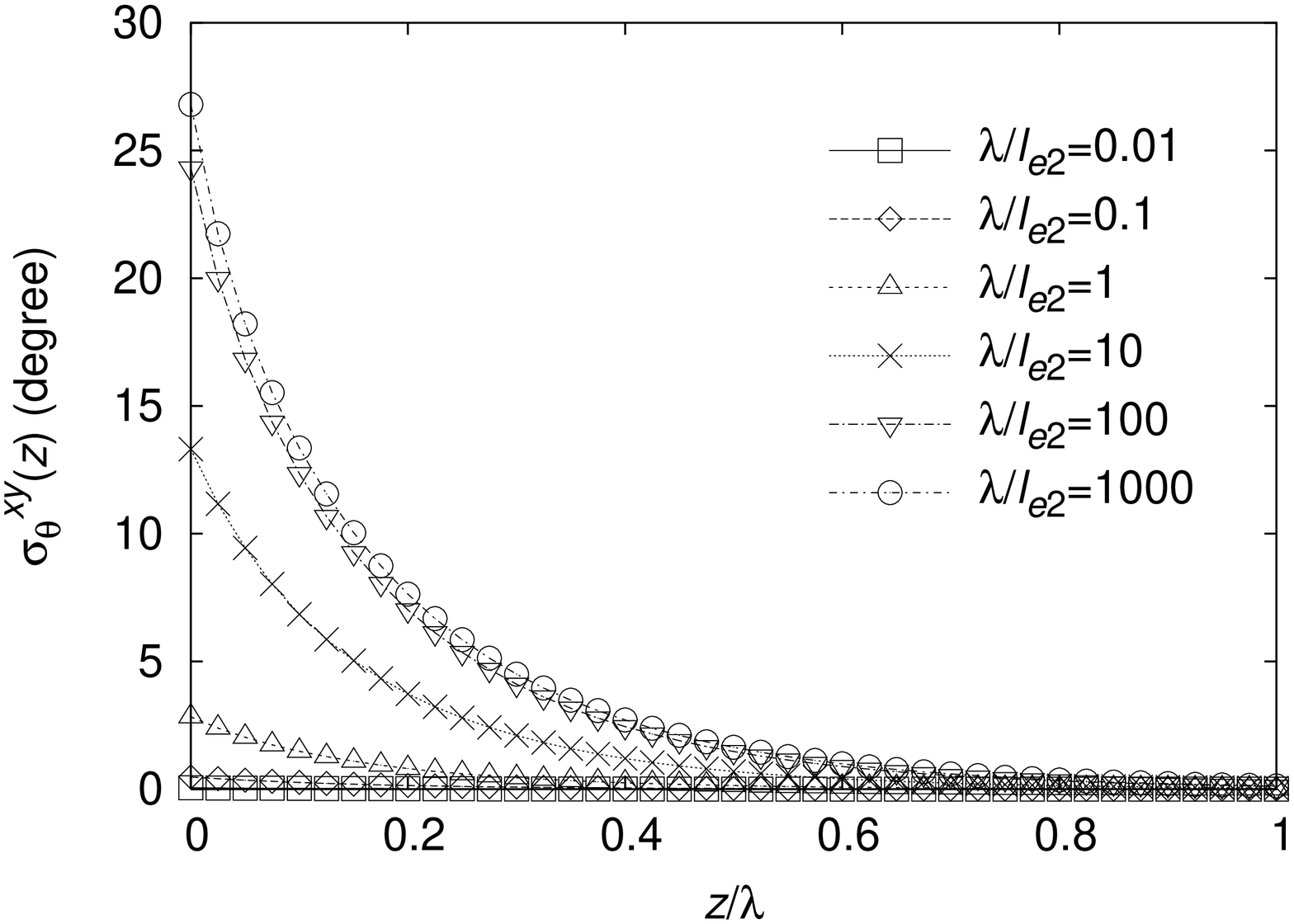}
\\
\\
\\
\\
\centerline{Fig.~\ref{sd_vs_depth} by Wan et al.}

\newpage
\includegraphics[height=\columnwidth,angle=270]{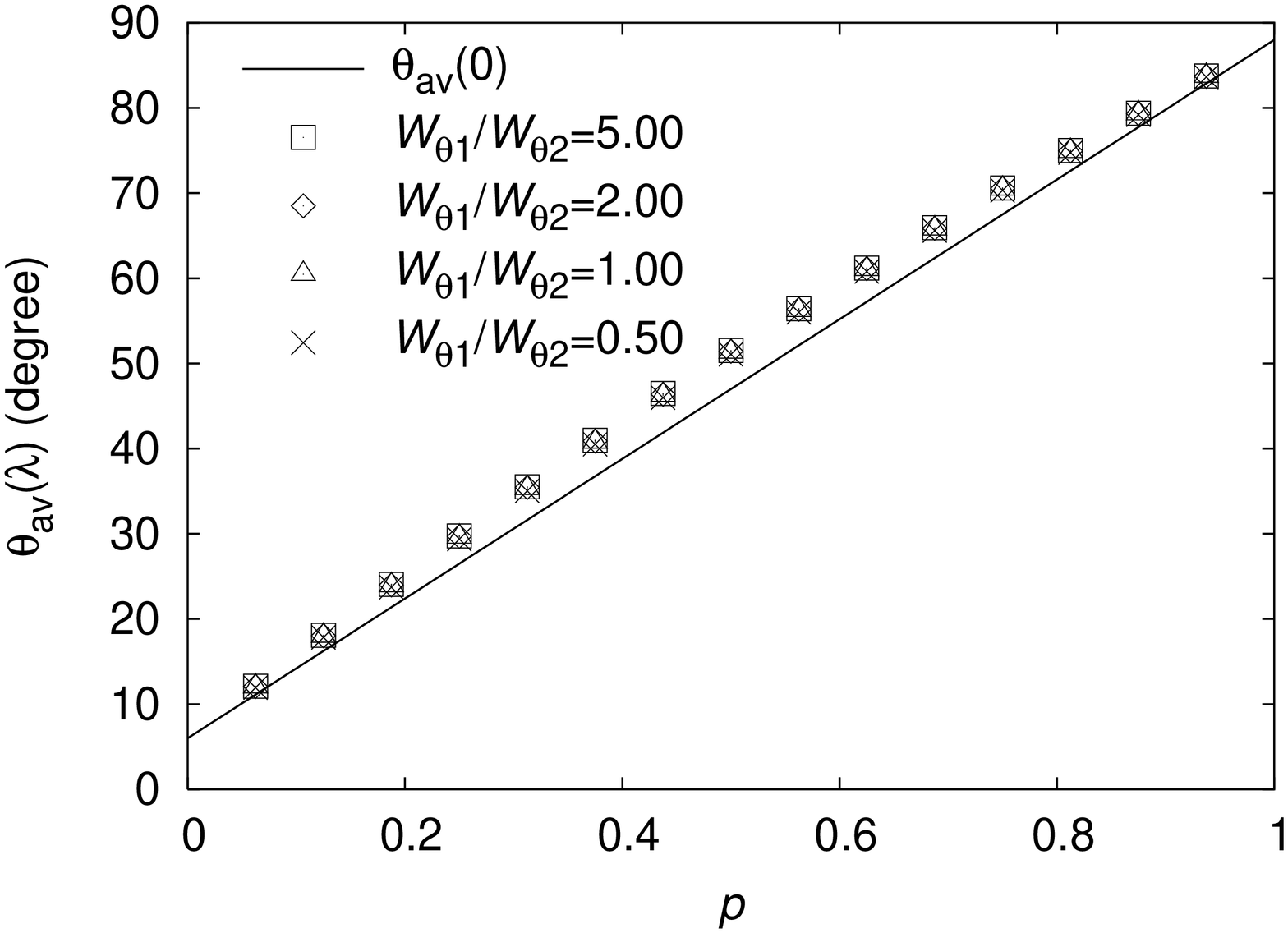}
\\
\\
\\
\\
\centerline{Fig.~\ref{pretilt_vs_area_lw_1000} by Wan et al.}

\end{document}